\documentclass[aps,pre,reprint,superscriptaddress,longbibliography,showpacs]{revtex4-2}
\usepackage{xcolor}

\usepackage[english]{babel}
\usepackage{graphicx}% Include figure files
\usepackage{dcolumn}% Align table columns on decimal point
\usepackage{bm}% bold math
\usepackage{verbatim} % For comments and other
\usepackage{amsmath}  % For math
\usepackage{amssymb}  % For more math
\usepackage{listings} % For source code
\usepackage[caption=false]{subfig}   % For subfigures
\usepackage{graphics}
\usepackage{latexsym,epsfig}
\usepackage{empheq}
\usepackage{mathtools}
\usepackage{adjustbox}
\usepackage{lipsum}
\usepackage{dsfont}

\newcommand{\orcidicon}[1]{\href{https://orcid.org/#1}{\includegraphics[height=\fontcharht\font`\B]{orcid.png}}} 

\usepackage{placeins} %This contains \FloatBarrier command.

\usepackage[scr=boondox]{mathalfa}

\usepackage[scr=boondox]{mathalfa}

\def\Im{\ensuremath{{\operatorname{Im}}}}

\usepackage{tikz,bm} % Useful for drawing plots
\usepackage{circuitikz}
\usepackage[colorlinks=true,linkcolor=blue,citecolor=blue,urlcolor=blue]{hyperref}

\begin{document}
\onecolumngrid
\title{Resonant level model coupled to a Sachdev-Ye-Kitaev bath}

\author{Anastasia Enckell}
\email{anastasia.enckell@theorie.physik.uni-goettingen.de}
\author{Stefan Kehrein}
\email{stefan.kehrein@theorie.physik.uni-goettingen.de}

\affiliation{
 Institute for Theoretical Physics, Georg-August-Universität Göttingen, Germany
}

\date{\today}

\begin{abstract}
We investigate the non-equilibrium dynamics of a resonant level model coupled to a strongly interacting electron bath modeled by a Sachdev-Ye-Kitaev (SYK) model. Different from the well-investigated case of a structureless non-interacting Fermi gas bath leading to a temperature-independent exponential decay of the impurity orbital occupation, we find a temperature-dependent oscillatory decay. We attribute this difference to the lack of quasiparticles in the SYK model, which is reflected in its singular density of states at the Fermi level. Our results are exact and can be obtained analytically by mapping to a suitably structured Fermi gas bath as an ancillary model for the SYK bath.
\end{abstract}

\maketitle

\section{Introduction}
The investigation of quantum impurities has a long and important tradition in solid state physics \cite{Hewson_1993}.
Paradigmatic models like the Kondo model and the Anderson impurity model have played a key role in understanding low temperature experiments on materials containing magnetic or non-magnetic impurities, and in developing theoretical methods like the numerical renormalization group, Bethe ansatz for impurity models, boundary conformal field theory, etc. In addition, the investigation of quantum impurities has paved the road to a better understanding of bulk strongly correlated materials like heavy fermions and Mott-Hubbard insulators, where quantum impurities can be thought of as the underlying building blocks \cite{RevModPhys.68.13}. 
Initially the focus of these works was on the equilibrium behavior like thermodynamic quantities and linear response properties (magnetic susceptibility, transport, etc.), while more recently the non-equilibrium behavior after time-dependent driving or beyond linear response transport have been much investigated.

Common to most of these studies is that the quantum impurity is coupled to a Fermi gas, that is a non-interacting quantum bath. Most analytical and numerical techniques rely on this feature because modeling an interacting fermionic bath is much more challenging, both analytically and numerically. However, it is expected that there are no qualitative differences when coupling to a Fermi liquid. The underlying reason for this is the existence of well-defined quasiparticles in a Fermi liquid, so one only expects quantitative changes due to the renormalized electron mass and density of states. This reasoning is confirmed by perturbative calculations \cite{PhysRevLett.70.4007}. 

In recent years there has been considerable interest in a new class of materials named strange metals, which cannot be described within the Fermi liquid paradigm \cite{PhysRevX.5.041025}. 
Strange metals show a linear in~$T$ resistivity behavior down to the lowest temperature, which is inconsistent with a quasiparticle description. This conclusion has been corroborated by Shot noise measurements \cite{doi:10.1126/science.abq6100}. So it is a very natural question to ask about the behavior of quantum impurities in such very strongly correlated electron systems where the Fermi liquid quasiparticle picture is not applicable.

Due to the strongly interacting nature of the bath electrons, this appears to be a formidable theoretical task. However, the Sachdev-Ye-Kitaev (SYK) model has given us a theoretical framework to study such non-quasiparticle fermionic models exactly in the thermodynamic limit using analytical techniques \cite{PhysRevLett.70.3339, kitaev, RevModPhys.94.035004}. 
This makes the SYK model an interesting bath model for quantum impurities, exploring a very different setting than the conventional non- or weakly interacting electron bath. 

A first step in this direction is looking at the simplest quantum impurity, namely the resonant level model (non-interacting Anderson impurity model) coupled to an SYK bath. In equilibrium this has been done in Ref.~\cite{PhysRevB.98.081413}, 
which studied tunneling spectroscopy in the SYK-model. The tunneling Hamiltonian investigated in Ref.~\cite{PhysRevB.98.081413} is just a sum over resonant level models. The main difference from a structureless Fermi gas comes from the local density of states. This is then reflected in the behavior of the differential conductance as a function of voltage bias. 

In this work we move beyond equilibrium and study the non-equilibrium dynamics of a resonant level model coupled to an SYK bath, specifically the decay of the impurity orbital occupation to its equilibrium value. We focus on a charge neutral SYK model with the impurity orbital energy aligned with the Fermi level, where differences to a structureless Fermi gas bath are most noticeable. Using exact analytical methods we calculate the decay of the impurity level occupation, which shows oscillatory behavior and is temperature dependent. This is in stark contrast to a structureless Fermi gas bath, where the decay is purely exponential without any temperature dependence \cite{Guinea85}. 
We interpret this different behavior as being due to the lack of low-energy quasiparticles in the SYK bath.

Our paper is structured as follows. In Sect.~II we briefly review the complex SYK model that will be used as a quantum bath for the resonant level model. In Sect.~III we then introduce the model investigated in this paper, namely the RLM coupled to this SYK bath, and derive its equilibrium properties. 
The subsequent Sect.~IV contains the derivation of the Kadanoff-Baym equations, which describe the non-equilibrium dynamics of an RLM coupled to either a Fermi gas or an SYK bath. We take advantage of this in Sect.~V by introducing an ancillary model, where a suitably structured Fermi gas bath models the SYK bath from the point of view of the impurity dynamics both in equilibrium and non-equilibrium. The analytical solution of this ancillary model is worked out in Sect.~VI, which then contains the central results of our paper, namely the non-equilibrium dynamics of the impurity orbital occupation for different $q$-body interacting SYK models at various temperatures. Sect.~VII contains our conclusions and an outlook.

\section{SYK Model}
\begin{figure*}[t]
    \centering
    % First subfigure
\includegraphics[width=0.9\textwidth]{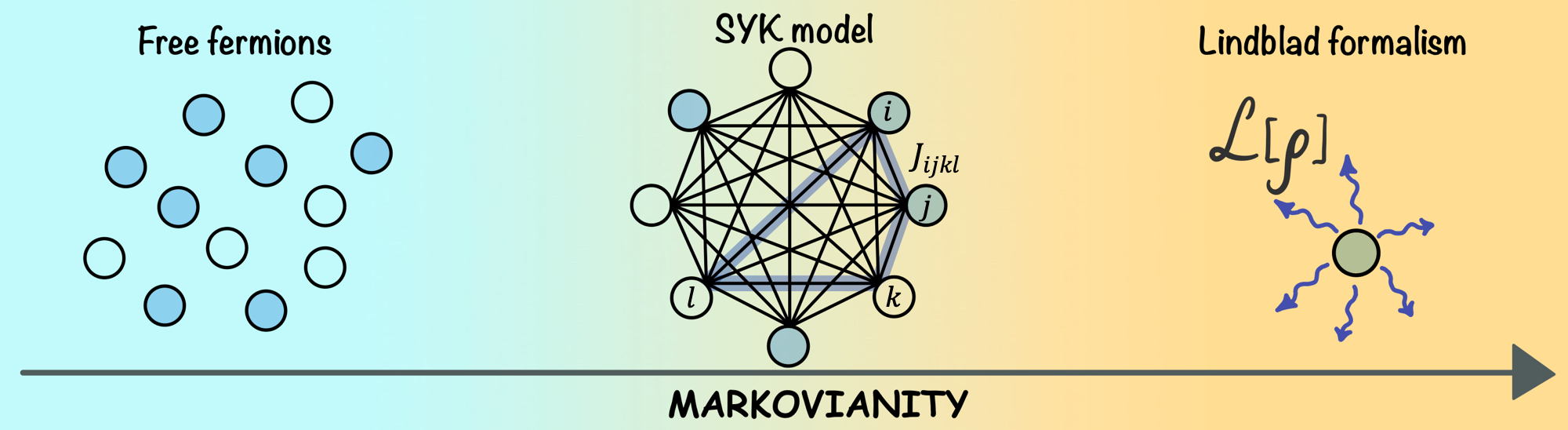}
    \caption{Comparison of different bath models.}
    \label{fig:different baths}
\end{figure*}

The complex SYK-$q$ model with \(q/2\)-body interactions is defined by the Hamiltonian \cite{PhysRevB.95.155131} 
\begin{equation}
    H_{SYK_q} = \sum\limits_{\substack{1\leq i_1< \hdots < i_{q/2}\leq N \\ 1\leq j_1< \hdots < j_{q/2}\leq N}}^N J^{j_1, \hdots, j_{q/2}}_{i_1,\hdots, i_{q/2}}c^\dagger_{i_1} \hdots c^\dagger_{i_{q/2}} c_{j_1} \hdots c_{j_{q/2}},
\end{equation}
where \(N\) is the number of lattice sites, \(c_k^\dagger\) and \(c_k\) are fermionic creation and annihilation operators and \(J_{i_1,\hdots, i_{q/2}; j_1, \hdots, j_{q/2} }\) are independent random variables drawn from a Gaussian distribution with zero mean and variance
\begin{equation}
    \overline{|J_{i_1,\hdots, i_{q/2}; j_1, \hdots, j_{q/2} }|^2} = \frac{J^2\left[(q/2)!\right]^2}{\left[N/2\right]^{q-1}},
\end{equation}
where \(J\) is a positive constant.  

The conserved quantity of this model is the total charge, which can be continuously varied with the chemical potential. We denote the charge density as 
\begin{equation} \label{charge density}
Q = \frac{1}{N} \sum_{k=1}^N \left \langle c_k^\dagger c_k\right \rangle - \frac{1}{2},
\end{equation}
such that $Q=0$ is the half-filled (particle-hole symmetric) point.
The simplest nontrivial case is \(q=4\) with two-body interactions is given by the Hamiltonian 
\begin{equation}
    H_{SYK_4}   = \sum^N_{1<i<j<k<l<N} J_{ijkl}c^\dagger_{i}c^\dagger_{j} c_{k} c_{l} \, ,
\end{equation}
where
\begin{equation}
    \overline{|J_{ijkl}|^2} = \frac{J^2}{(N/2)^{3}} \ .
\end{equation}
The SYK-$q$ models with $q=4,6,8,\ldots$ behave very similarly with only quantitative differences \cite{Gross_2017}.  \\ 
A detailed derivation of the saddle point equations for SYK-4 and generalizations to SYK-\(q\) models were discussed in Refs.~\cite{PhysRevX.5.041025,RevModPhys.94.035004}.
The saddle point equations for large-N are
\begin{equation}
\begin{aligned} \label{syk kb}
        G(i\omega_n) &= \frac{1}{i\omega_n- \Sigma(i\omega_n)} ,\\
    \Sigma(\tau) &= -J^2G^{q/2}(\tau)G^{q/2-1}(-\tau),
\end{aligned}
\end{equation}
where the Euclidean propagator is defined as
\begin{equation}
    G(\tau_1, \tau_2) = -\frac{1}{N} \sum_i\left \langle T\left (c_i(\tau) c^\dagger_i(\tau_2) \right) \right \rangle .
\end{equation}
The SYK model is exactly solvable for \(1 \ll J\beta  \ll N\), where the equations (\ref{syk kb}) are invariant under conformal reparametrisation. In the energy range \(J/N \ll \omega \ll J\), the solution for the zero temperature Green's function is \cite{Gu:2019jub}
\begin{equation} \label{green's zero}
    G(\pm i\omega) = \mp i e^{\mp i \alpha} \sqrt{\frac{\Gamma(2-2/q)}{\Gamma(2/q)}}b^{(1/q-1/2)}J^{-q/2}\, \omega^{(2/q-1)} .
\end{equation}
Here
\begin{equation}
    b= \frac{1-2/q}{2\pi}\frac{\sin{(2\pi/q)}}{2\cos{\left(\pi(1/q+i\mathcal{E})\right)}\cos{(\pi(1/q-i\mathcal{E}}))},
\end{equation}
with the asymmetry angle  $\alpha \in \left(-\pi/q, \pi/q \right)$ and the relation
\begin{equation}
    e^{2 \pi \mathcal{E} } = \frac{\sin \left(\pi/q+\alpha\right)}{\sin \left(\pi/q-\alpha\right)}.
\end{equation}
The asymmetry angle relates to the charge density in eq. (\ref {charge density}) via 
\begin{equation}
Q = -\frac{\alpha}{\pi}-\left(\frac{1}{2}-\frac{1}{q}\right) \frac{\sin (2 \alpha)}{\sin{(2\pi/q)}} ,
\label{theta and Q}
\end{equation}
In the sequel we will only be interested in the half-filled case, therefore
$\alpha=0$ leading to $\mathcal{E}=0$ and
\begin{equation}
b=\frac{1-2/q}{2\pi}\:\tan(\pi/q) \ .
\end{equation}
Analytically continuing to the real frequencies we find
\begin{eqnarray}\label{syk zero T}
    G^R(\omega) =&& -ie^{-\text{sgn}(\omega)i\pi(2/q-1)/2}\sqrt{\frac{\Gamma(2-2/q)}{\Gamma(2/q)}} \nonumber \\
    && \times b^{(1/q-1/2)}J^{-2/q}|\omega|^{(2/q-1)}.
\end{eqnarray}
At non-zero temperature the conformal solution is \cite{Gu:2019jub} 
\begin{eqnarray}\label{syk nonzero T}
    G^R(\omega) =&& -  i  \sqrt{\frac{\Gamma(2-2/q)}{\Gamma(2/q)}}J^{-2/q}\left(2\pi T \sqrt{b}\right)^{(2/q-1)} \nonumber \\
    && \times \frac{\Gamma\left(\frac{-i\omega}{2\pi T } + \frac{1}{q} \right)}{\Gamma\left(\frac{-i\omega}{2\pi T } +1 - \frac{1}{q} \right)}.
\end{eqnarray}
The small-$\omega$ power law behavior of the Green's function in eq. (\ref{green's zero}) indicates non-Fermi liquid behavior without an underlying quasiparticle picture. Furthermore, the SYK-model is non-integrable and even maximally chaotic with a Lyapunov exponent saturating the Maldacena-Shenker-Stanford bound \cite{PhysRevD.94.106002}, indicating rapid information scrambling. These properties are consistent with the observation that the SYK-model satisfies the Eigenstate Thermalization Hypothesis (ETH) \cite{Sonner:2017hxc}.
Moreover, the SYK-model has a holographic connection to the AdS\(_2\) horizon of a charged black hole.
In summary, the SYK model captures important aspects of many-body quantum chaos, quantum thermalization and information scrambling while being analytically solvable. This has led to enormous interest in its properties in the past decade from the viewpoints of condensed matter physics, quantum information theory and fundamental physics \cite{PhysRevB.95.155131, PhysRevB.97.245126,PhysRevB.98.125134, PhysRevX.8.031024}. 

In this paper we are exploring a new direction to take advantage of the unique properties of the SYK model, namely to use it as a quantum bath. Generally speaking, a quantum bath models the environment of some quantum system in order to describe environment-induced features like decoherence and dissipation \cite{Weiss:2021uhm}. 
There are two main theoretical approaches for this purpose, namely free particle baths like harmonic oscillators 
(see e.g. Ref.~\cite{RevModPhys.59.1})  
or the Lindblad formalism \cite{Breuer2002}.  
These two approaches can be considered opposing limits for modeling an environment: Non-interacting free particle baths are integrable, do not obey ETH and do not cause scrambling. On the other hand the Lindblad approach applies a Markovian approximation, which eliminates all quantum coherence effects in the environment. While both of these approaches are perfectly suitable for certain situations, there are clearly situations where it would be desirable to model an environment that both obeys ETH and has non-trivial quantum coherence properties. One such situation is e.g. a non-zero temperature environment where the quantum system explores the ETH-part of the environment spectrum, while still retaining some quantum coherence in the environment. From the theoretical point of view this is a challenging setup, because an interacting bath will usually require analytical approximations or considerable numerical resources. However, the SYK model is analytically solvable and 
situated somewhere between free particle bath and the Lindblad approach (see Fig.~\ref{fig:different baths}), 
which makes it an interesting and natural candidate for a quantum bath. 

\section{Resonant level coupled to SYK bath}
\begin{figure*}[t]
        \centering
        \subfloat[]{\includegraphics[width=0.45\linewidth]{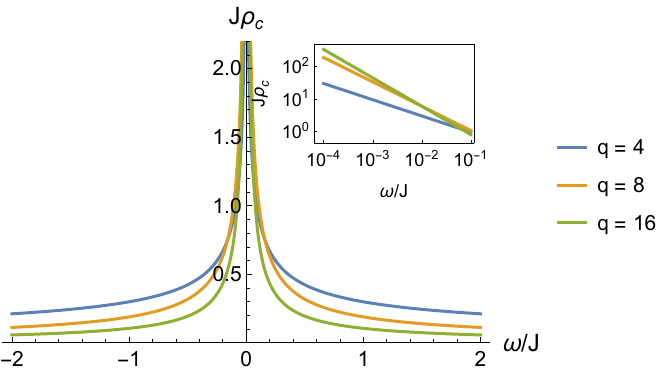}}
        \hfill
        \subfloat[]{\includegraphics[width=0.45\linewidth]{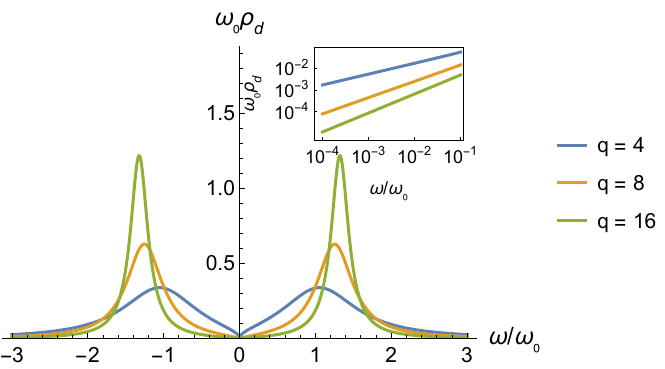}}\qquad
         \subfloat[]{\includegraphics[width=0.45\linewidth]{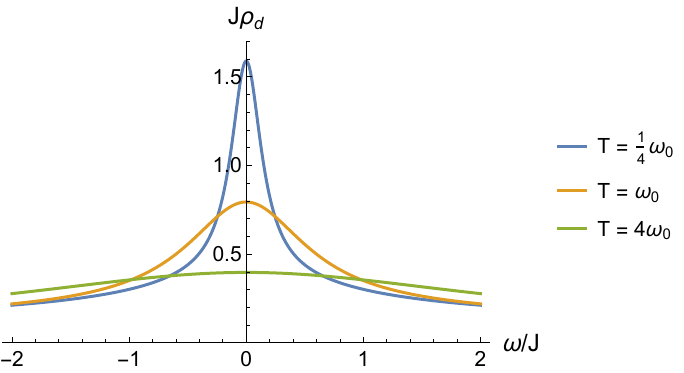}}
        \hfill
        \subfloat[]{\includegraphics[width=0.45\linewidth]{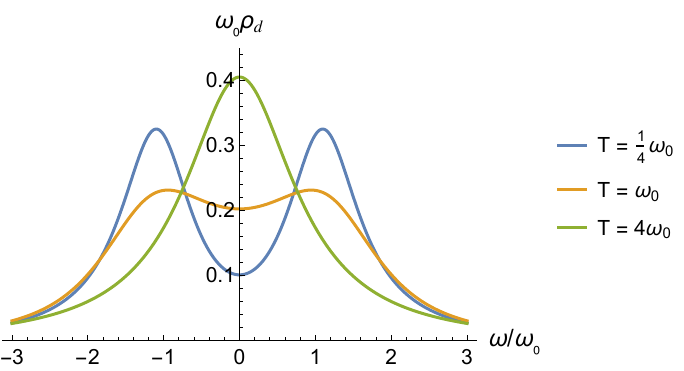}}
% \captionsetup{justification=centering}
\caption{Top row: Density of states of the SYK bath $\rho_c(\omega)$ (left) and  the impurity $\rho_d(\omega)$ for \(q = 4, 8, 16\) (right) for zero temperature. Notice the divergence $\propto |\omega|^{-1+2/q}$ for the SYK bath and the non-analytic suppression $\propto |\omega|^{1-2/q}$ for the local impurity density of states at the Fermi energy. Bottom row: Density of states of the SYK-4 bath (left) and the impurity (right) for non-zero temperature. Notice that the behavior at the Fermi energy becomes analytic for non-zero temperature. The plots are scaled with the characteristic energy scales \(J\) and \(\omega_0\) (eq. (\ref{scaling}))  for the SYK bath and the impurity, respectively, and show the universal behavior as a function of these energy scales.}
\label{fig:DOS bath and impurity}
\end{figure*} 
Motivated by this 
we consider the following Hamiltonian, which describes a non-interacting impurity coupled to an SYK bath 
\begin{equation}\label{hamiltonian}
    H_\text{RLM-SYK} = \epsilon_d d^\dagger d + H_\text{SYK} + \frac{V}{\sqrt{N}}\sum\limits_{i=1}^N  \left(c_i^{\dagger} d+ d^{\dagger} c_i\right).
\end{equation}
The impurity with energy \(\epsilon_d\) is connected to each of the \(N\)  sites of the SYK bath via the constant hybridization \(V\). If one replaces $H_\text{RLM-SYK}$ in eq. (\ref{hamiltonian}) with a non-interacting Fermi gas,
one obtains the conventional and well-studied resonant level model (non-interacting Anderson impurity model) \cite{Hewson_1993} 
\begin{equation}
\label{Hamiltonian_RLM_FG}
H_\text{RLM-FG}=\epsilon_d d^\dagger d + \sum_{i=1}^N \epsilon_i c^\dagger_i c_i + \frac{V}{\sqrt{N}}\sum\limits_{i=1}^N  \left(c_i^{\dagger} d+ d^{\dagger} c_i\right).
\end{equation}
Since the impurity is non-interacting we can consider spinless fermions in both equations (\ref{hamiltonian}) and (\ref{Hamiltonian_RLM_FG}). 

We will be interested in the time-dependent occupation of the impurity after instantaneous coupling to the bath. The key question we are addressing is how the impurity dynamics 
is affected by replacing the non-interacting Fermi gas bath in the conventional resonant level model  
(\ref{Hamiltonian_RLM_FG}) by the strongly interacting, non-integrable and chaotic SYK bath (\ref{hamiltonian}). From the experimental point of view this corresponds to studying a non-magnetic impurity in a strange metal host, where the strange metal is modeled with an SYK Hamiltonian.   

The equilibrium Dyson equations for this setup can be derived from the path integral or diagrammatic expansion \cite{PhysRevX.5.041025,PhysRevD.94.106002}. 
 Decoupling of the interaction term with consecutive variation with respect to bilocal fields does not influence the impurity part of the action, therefore its contribution can be found separately \cite{Zhong_2018}.
Denoting the impurity with an index \(d\) and the bath with an index \(c\), one finds the following Dyson equations in the large-$N$ limit
\begin{subequations}
\begin{align}
    G_c(i\omega_n) &= \frac{1}{i\omega_n - \Sigma_c(i\omega_n)} \label{bath gf},\\
    \Sigma_c(\tau) &= -J^2G_c^{q/2}(\tau)G_c^{q/2-1}(-\tau) + \frac{1}{N}V^2G_d(\tau) \label{bath self e},\\
    G_d(i\omega_n) &= \frac{1}{i\omega_n +\epsilon_d- \Sigma_d(i\omega_n)} \label{imp gf},\\
    \Sigma_d(\tau) &= V^2 G_c(\tau) ,\label{imp self energy}
\end{align}
\end{subequations}
where \(\omega_n\) is a Matsubara frequency and \(\tau\) is imaginary time (for details of the calculation see Appendix~A).
As \(N\to \infty\), the impurity contribution to the bath self-energy vanishes, making the bath dynamics independent of the impurity. For the same reason the change of the bath chemical potential can be neglected.   \\
After analytical continuation to retarded Green's functions, we can obtain the equilibrium density of states of the bath and the impurity from the imaginary part of the Green's function
\begin{equation} \label{rho}
\begin{split}
        \rho_c(\omega) &= - \frac{1}{\pi}\Im \: G_c^R(\omega) \ , \\
        \rho_d(\omega) &= - \frac{1}{\pi}\Im\: G_d^R(\omega) \ . 
\end{split}
\end{equation}
Due to the negligible impurity influence in the $N\rightarrow\infty$ limit,  \(\rho_c(\omega) \) is given by eq. (\ref{syk zero T}) and eq. (\ref{syk nonzero T}) for zero and non-zero temperature, respectively. For zero temperature this implies the well-known non-analytic behavior of the SYK density of states at the Fermi energy,
\begin{equation}
\rho_c(\omega)\propto |\omega|^{-1+2/q} \ .
\label{eq_rhoc}
\end{equation}
Specifically for $q=4$,
\begin{equation}
\rho_c(\omega)=\frac{1}{2^{1/2}\pi^{3/4}}\,|J\,\omega|^{-1/2} \ .
\end{equation}

In order to find the impurity density of states \(\rho_d(\omega)\), we first split the self energy into real and imaginary part,
\begin{equation}
    \Sigma^R_d(\omega) = \Lambda(\omega) + i \Delta(\omega),
\end{equation}
which are related by Kramers-Kronig relations
\begin{subequations}
\begin{align} \label{real self energy}
    \Delta(\omega) &=  -\mathcal{P} \int \frac{d \omega'}{\pi} \frac{\Lambda(\omega')}{\omega' - \omega}, \\ \label{imag self energy}
    \Lambda(\omega) &= \mathcal{P} \int \frac{d \omega'}{\pi} \frac{\Delta(\omega')}{\omega' - \omega} .
\end{align}
\end{subequations}
According to eq. (\ref{imp self energy}), \(\Sigma_d^R(\omega)\) is identified with the 
hybridization function of the resonant level model, 
\begin{equation}\label{delta}
    \Delta(\omega) = \pi\,V^2\,\rho_c(\omega) , 
\end{equation}
which describes the broadening of the impurity energy level due to the hybridization with a bath.

One can easily verify that the hybridization function is a universal function of the energy scale
\begin{equation}
\omega_0=V\,\left(\frac{V}{J}\right)^{1/(q-1)} \ ,
\label{scaling}
\end{equation}
namely
\begin{equation}
    \Delta(\omega)=\omega_0\,\tilde\Delta(\omega/\omega_0,T/\omega_0) \ .
\end{equation}
Here $\tilde\Delta(\tilde\omega,\tilde T)$ is a dimensionless function in terms of 
the dimensionless variables
\begin{equation}
\tilde\omega=\frac{\omega}{\omega_0} \quad ,\quad
\tilde T=\frac{T}{\omega_0} \ .
\end{equation}
For $T=0$, one has
\begin{equation}
\tilde\Delta(\tilde\omega,0)\propto |\tilde\omega|^{-1+2/q}
\end{equation}
and at nonzero temperature
\begin{equation}
\tilde\Delta(\tilde\omega,\tilde T) \propto
    {\tilde T}^{-1+2/q} \frac{\Gamma\left(\frac{-i\tilde\omega}{2\pi \tilde T } + \frac{1}{q}\right)}{\Gamma\left(\frac{-i\tilde\omega}{2\pi \tilde T } +1 - \frac{1}{q}\right)}.
\end{equation}
Specifically for $q=4, T=0$,
\begin{equation}
\tilde\Delta(\tilde\omega)=\frac{\pi^{1/4}}{2^{1/2}} \,|\tilde\omega|^{-1/2} \ .
\end{equation}
Substituting the self-energy back into the Green's function, we get
\begin{equation}
    G^R_d(\omega) = \frac{(\omega -\epsilon_d-\Lambda(\omega))+i\Delta(\omega)}{(\omega -\epsilon_d-\Lambda(\omega))^2+\Delta^2(\omega)} .
    \label{eq_Gd_equ}
\end{equation}
Taking the impurity energy to be at the Fermi surface, \(\epsilon_d = 0\), the equilibrium density of states of the impurity becomes
\begin{eqnarray}
\rho_d(\omega) &=& - \frac{1}{\pi}\Im( G^R_d\left(\omega)\right) \nonumber \\
&=&  \frac{1}{\pi}\frac{\Delta(\omega)}{\left(\omega - \Lambda(\omega)\right)^2 + \Delta^2(\omega)} \nonumber \\
&=&\frac{1}{\omega_0}\,\frac{1}{\pi}\,
\frac{\tilde\Delta(\tilde\omega,\tilde T)}{
(\tilde\omega-\tilde\Lambda(\tilde\omega,\tilde T))^2+
\tilde\Delta^2(\tilde\omega,\tilde T)
} \ ,
\label{impurityDOS}
\end{eqnarray}
that is a universal function of the energy scale~$\omega_0$. One easily verifies, that
\begin{equation}
\tilde\Lambda(\tilde\omega,0) \propto |\tilde\omega|^{-1+2/q},
\end{equation}
and, therefore, at zero temperature,
\begin{equation}
\omega_0\,\rho_d(\omega) \propto |\tilde\omega|^{1-2/q} ,
\end{equation}
indicating a non-analytic power law suppression of the impurity density of states at the Fermi level for $q=4,6,\ldots$.
Specifically for $q=4$, $T=0$,
\begin{equation}
\omega_0\,\rho_d(\omega)=\frac{2^{1/2}}{\pi^{5/4}}\,
\frac{|\tilde\omega|^{1/2}}{1+\left(1-\frac{2^{1/2}}{\pi^{1/4}}|\tilde\omega|^{3/2}\right)^2} ,
\end{equation}
with a square root non-analytic behavior at the Fermi energy. At non-zero temperature the non-analyticity disappears and one finds
\begin{equation}
\omega_0\,\rho_d(0)=\tilde T^{1-2/q} \ .
\end{equation}

Examples for the SYK bath density of states and the local impurity density of states at zero and non-zero temperature are shown in Fig.~\ref{fig:DOS bath and impurity}. 

%-----------------------------------------------------

\section{Kadanoff-Baym equations}
The above equilibrium result for the local impurity density of states (\ref{eq_Gd_equ}) only depends on the bath density of states irrespective of whether the bath is interacting or not. This is a well known property of non-interacting quantum impurity models like the resonant level model discussed here \cite{Hewson_1993}. It is not immediately obvious that this property also extends to non-equilibrium situations, therefore we will briefly go through the necessary steps to verify this explicitly.

To proceed to the non-equilibrium case, we consider the real time Green's function 
\begin{equation}
    G_d(t,t') = -\left \langle T_c \{d^\dagger(t') d(t) \}\right \rangle,
\end{equation}
where \(T_c\) is a contour time-ordering operator defined on a Keldysh contour.
We also define the lesser, greater, as well as retarded and advanced Green's functions on the contour as follows
\begin{subequations}
\begin{align}
G_d^<(t, t')  &= G_d(t_{-}, t'_{+}) = i \left \langle d^\dagger(t') d(t)\right \rangle, \\
     G_d^>(t, t')  &= G_d(t_{+}, t'_{-}) = -i  \left \langle  d(t) d^\dagger(t')\right \rangle, \\
    G_d^R(t,t') &= -\theta(t-t') \left(G_d^>(t,t') -G_d^<(t,t')\right),\\
G_d^A(t,t') &= \theta(t'-t) \left(G_d^<(t, t') -G_d^>(t,t')\right) .
\end{align}
\end{subequations}
where \(t_{-}\) refers to the time on the lower part and \(t_{+}\) refers to the upper part of the contour.
The general form of the non-equilibrium Dyson equations read
\begin{align}\label{dyson1}
    (i\partial_t-\epsilon_d)G_d(t,t') &= \delta_C(t-t') + \int_C ds \, \Sigma_d(t,s)G_d(s,t'), \\\label{dyson2}
    (-i\partial_{t'}-\epsilon_d) G_d(t,t') &= \delta_C(t-t') + \int_C ds\,  G_d(s,t')\Sigma_d(t,s).
\end{align}
Applying Langreth rules to these equations and using the above definitions of the retarded and advanced functions, we get the Kadanoff-Baym equations in a form 
\begin{widetext}
\begin{align}
 (i\partial_t -\epsilon_d)G^\lessgtr_d(t,t') &= \int_{-\infty}^t ds \,\Sigma_d^R(t,s) G^\lessgtr_d(s,t') + \int_{-\infty}^{t'} ds \,\Sigma_d^\lessgtr(t, s) G_d^A(s,t'), \\
 (-i\partial_{t'} -\epsilon_d)G_d^\lessgtr(t,t') &= \int_{-\infty}^t ds\, G_d^R(t,s) \Sigma_d^\lessgtr(s,t') + \int_{-\infty}^{t'} ds\, G_d^\lessgtr(t, s) \Sigma_d^A(s,t').
\end{align}
\end{widetext}
Due to
\begin{equation}
\Sigma_d^\lessgtr(t,s)=V^2\,G_c^{\lessgtr}(t,s)
\end{equation}
the impurity self-energy deviates from the equilibrium result in order \(\frac{1}{N}\), which becomes negligible in the thermodynamic limit. Therefore, the non-equilibrium impurity Green's function \(G_d\) is fully determined by the equilibrium bath Green's function \(G_c\). Just like in the equilibrium situation discussed above, only the bath density of states matters irrespective of whether the bath is interacting or not.

\section{Analytical solution}
In order to take advantage of the above observation, we now review the occupation dynamics of an initially fully occupied impurity site coupled to a Fermi gas with an arbitrary density of states. 
We consider the Hamiltonian
\begin{equation}
H=\epsilon_d d^{\dagger} d+ \sum\limits_{k=1}^N \epsilon_k c^\dagger_k c_k + \frac{1}{\sqrt{N}}\sum\limits_{k=1}^N V_k\,( c_k^{\dagger} d+d^{\dagger} c_k)
\end{equation}
with $\epsilon_d=\epsilon_F=0$ and overall particle-hole symmetry like in Fig.~\ref{fig:DOS bath and impurity} for the SYK bath.  
We calculate the occupation of the impurity orbital as a function of time 
\begin{equation} \label{nd}
    n_d(t)  = \left \langle d^\dagger(t) d(t) \right \rangle . 
\end{equation}
After equilibration the impurity site will be half filled,
\begin{equation}
\lim_{t\rightarrow\infty} n_d(t)=1/2 \ .
\end{equation}
The dynamics of the time dependent impurity creation operator \(d(t)\) is governed by the Heisenberg equation of motion
\begin{equation}
    \frac{d }{dt} d(t) = \frac{i}{\hbar}[H, d(t)],
\end{equation}
which, for the case of a free Fermi bath, has a general solution in a form 
\begin{equation}\label{ansatz}
    d(t) = h(t)\, d + \sum_k \gamma_k(t)\, c_k , 
\end{equation}
where
\begin{equation}\label{h function}
    h(t) = \left \langle \{ d(t) , d^\dagger(0)\}\right \rangle.
\end{equation}
Combining this with eq. (\ref{nd}), we get 
\begin{equation}
    n_d(t) =h^2(t)\left \langle d^\dagger(0) d(0)\right \rangle + \sum_k \gamma_k^2(t)\langle c^\dagger_k(0) c_k(0)\rangle \ .
\end{equation}
Since the impurity site is initially occupied and $\epsilon_F=0$ this becomes
\begin{equation} \label{occuppation half filled}
    n_d(t) = h^2(t) + \sum_{\epsilon_k <0}
    \gamma_k^2(t).
\end{equation}
Notice that this result also holds for non-zero bath temperature since the coefficients $\gamma_k(t)$ are symmetric around the Fermi energy.
As a matter of fact $h(t)$ is already determined by the equilibrium result for the impurity orbital density of states. One can see this from 
\begin{eqnarray}
   G_d^R(t) &=& -i \theta(t) \left \langle \{ d(t) , d^\dagger(0)\}\right \rangle \nonumber \\
   &=& -i \theta(t) h(t) \ ,
\end{eqnarray}
where we have used eq. (\ref{h function}).
For positive times the spectral decomposition of the Green's function reads \cite{Mattuck:1976xt}
\begin{equation}
    G^R_d(t) = -i\int_{-\infty}^{\infty} d\omega e^{-it\omega} \rho_d(\omega) 
\end{equation}
and hence for a symmetric density of states,
\begin{equation}
\label{eq_ht}
    h(t) = \int_{-\infty}^{\infty} d\omega \cos(t\omega)  \rho_d(\omega) \quad  , \quad t>0 \ .
\end{equation}
To finalize the calculation we need the missing contribution involving the $\gamma_k(t)$ in eq. (\ref{occuppation half filled}). Interestingly, it is also determined by $h(t)$. We take advantage of the
fermionic anticommutation relation evaluated with the ansatz (\ref{ansatz}) 
\begin{eqnarray}
   1&=& \{d^\dagger(0), d(0) \} =\{d^\dagger(t), d(t)\} \nonumber \\
   &=& h^2(t) + \sum_k \gamma^2_k(t) \ .
\end{eqnarray}
Because of particle-hole symmetry,
\begin{equation}
    \sum_{ \epsilon_k <0} \gamma^2_k(t) = \frac{1}{2} \sum_k \gamma^2_k(t) = \frac{1}{2}\left(1-h^2(t)\right).
\end{equation}
According to eq. (\ref{occuppation half filled}) the impurity occupation as a function of time $t\geq 0$ is therefore given by
\begin{eqnarray}
    n_d(t) &=& h^2(t) + \frac{1}{2}\left(1-h^2(t)\right)
   \nonumber \\ 
    &=& \frac{1}{2}\left(1+h^2(t)\right) 
    \label{eq_nd} . 
\end{eqnarray}
with $h(t)$ being the Fourier transform (\ref{eq_ht}) of the equilibrium impurity density of states. Notice that in the particle-hole symmetric case the impurity occupation can never dip below half filling,
\begin{equation}
\forall t\quad n_d(t)\geq 1/2 \ ,
\end{equation}
and is independent of the bath temperature. 

%-----------------------------------------------------
\section{Ancillary model}

In the limit of large \(N\) the bath self-energy in eq. (\ref{bath self e}) becomes independent of the impurity. Following the previous section's analysis, we see that the impurity self-energy 
is therefore determined by \(\Delta(\omega) \) in eq. (\ref{delta}) via equations (\ref{imp self energy}) and (\ref{rho}).
This makes the solution of the Kadanoff-Baym equations
for the impurity Green's function only dependent on the quantity \(\pi V^2 \rho_c(\omega)\). 
Since we know \(\Delta(\omega)\) for the SYK bath, any setup with the same \(\Delta(\omega) \) will result in the same Kadanoff-Baym equations for the impurity Green's function. 
\begin{figure}[t]
   \centering
    % First subfigure
     \subfloat[]{
\includegraphics[width=0.48\textwidth]{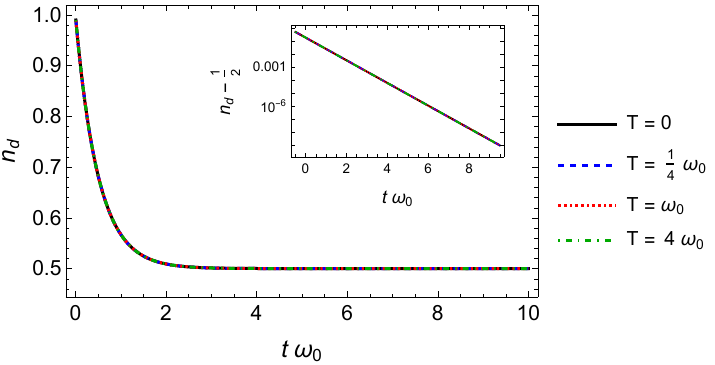}}
        \label{fig:subfig1}
    \hfill
    \subfloat[]{
\includegraphics[width=0.48\textwidth]{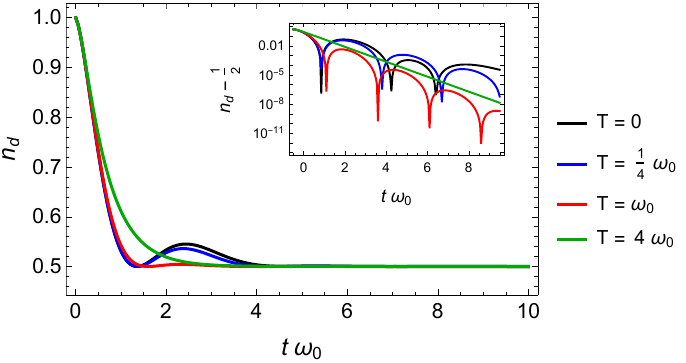}}
        \label{fig:subfig2}
    \hfill
    % Second subfigure
    \subfloat[]{
\includegraphics[width=0.48\textwidth]{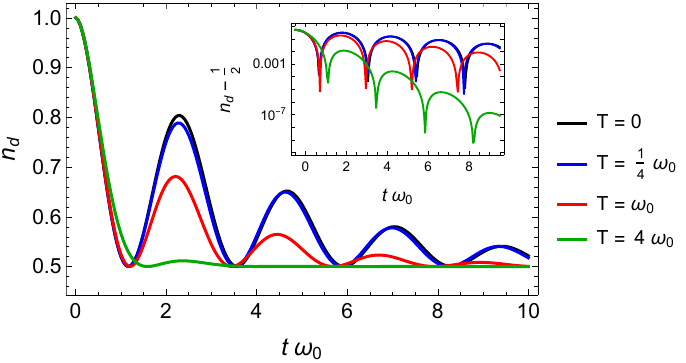}}
        \label{fig:subfig3}
    \caption{Impurity occupation as a function of time after a quench coupling to (a) SYK-2 bath, (b) SYK-4 bath, (c) SYK-16 bath. All curves are universal functions of the energy scale $\omega_0$ from (\ref{scaling}).}
    \label{fig:impurity analyt}
\end{figure}

We will apply this observation by considering an ancillary setup of an impurity coupled to a free Fermi bath with the same density of states as the SYK bath
\begin{equation} 
\label{rhocFG}
    \rho_{c,{\rm FG}}(\omega) =  \rho_c(\omega) \ .
\end{equation}
As shown above this yields the same impurity dynamics as the original SYK bath. (Notice that
alternatively we could have used a Fermi bath with a constant density of states and hybridization matrix elements with an appropriate energy dependence.) Now for the free Fermi bath we can use the analytical solution from Sect.~V, which we have derived for an arbitrary Fermi bath density of states. This is our key observation and we get our final result in eq. (\ref{eq_nd}) for the impurity occupation $n_d(t)$ with $h(t)$ being the Fourier transform (\ref{eq_ht}) of the impurity equilibrium density of states in eq. (\ref{impurityDOS}). While one cannot perform this Fourier transform in closed form analytically, it can be done numerically to arbitrary accuracy, as used in the following plots. One immediate consequence of the above relations is that $n_d(t)$ is a $q$-dependent universal function of the dimensionless parameters $t\,\omega_0$, $T/\omega_0$ with \(\omega_0\) from eq. (\ref{scaling}).

Figure \ref{fig:impurity analyt} shows the occupation of the impurity as a function of time for different values of~$q$ and different bath temperatures. The case $q=2$ corresponds to the conventional non-interacting Fermi gas bath with a flat band and one finds the well-known temperature-independent exponential decay towards the asymptotic value~$1/2$ \cite{Guinea85}. For both $q=4$ and $q=16$, that is genuinely interacting baths, one observes oscillatory behavior that becomes more pronounced for larger~$q$. In addition, now there is a temperature dependence with larger temperature suppressing the oscillations and leading to an exponential decay. 

This markedly different behavior of the interacting SYK bath with respect to oscillations and temperature dependence as compared to the flat band Fermi gas bath is the key result of this paper. Notice that the oscillations correspond to times~$t^c_i$ with $h(t^c_i)=0$ leading to $n_d(t^c_i)=1/2$. Careful numerical evaluation of the Fourier transform (\ref{eq_ht}) shows that there are 3~such times for~$q=4, T=0$, and 33 such times for $q=16, T=0$. This number increases for even larger values of~$q$. What happens at these times is that the excess impurity orbital occupation (meaning impurity occupation larger than the asymptotic value~$1/2$) has decayed completely into the bath, but then subsequently returns with a reduced amplitude to the impurity. This behavior is surprising from the point of view of the original strongly interacting SYK bath as one might have expected that it is impossible for the electron to return to its original state after having been completely absorbed by the interacting bath. Notice that similar oscillatory behavior has been found in numerical studies of the quasiparticle residue for an SYK model coupled to non-interacting leads in Ref.~\cite{PhysRevResearch.2.013307}, whereas the study of density inhomogeneity in a many-body pure state in the SYK model in Ref.~\cite{9cbp-7tvv} shows an opposing trend: oscillations appear in the SYK-2 model but not in the SYK-4 model.  

From the above analysis it is clear that the oscillatory behavior of $n_d(t)$ is due to the singular behavior of the SYK bath density in eq. (\ref{eq_rhoc}) at the Fermi energy. This in turn is related to the absence of quasiparticles in the SYK model, which lies at the heart of the differences between structureless Fermi gas bath and SYK bath considered here. 

\section{Conclusions}
In this paper we presented an analytical solution of the problem of a non-interacting impurity  coupled to an SYK bath (eq. (\ref{hamiltonian})). We considered the impurity at Fermi energy and the bath to be half-filled. We showed that in the thermodynamic limit the behavior of the impurity can be be obtained from an
ancillary model with a non-interacting Fermi bath and a suitably chosen density of states in eq. (\ref{rhocFG}), both in equilibrium and in non-equilibrium. Equilibrium spectral functions (Fig.~\ref{fig:DOS bath and impurity}) and non-equilibrium occupation decay $n_d(t)$ (Fig.~\ref{fig:impurity analyt}) are universal functions of the energy scale $\omega_0$ (\ref{scaling}), that is they only depend on dimensionless time $\tilde t=t\,\omega_0$ and temperature $\tilde T=T/\omega_0$. Interestingly one observes oscillatory behavior in $n_d(t)$ for $q=4,6,\ldots$ at low enough temperatures, implying that the interacting bath repeatedly regurgitates an electron back to the impurity orbital. These effects become more pronounced for larger~$q$. The oscillatory behavior disappears at larger temperature $\tilde T\gg 1$, where one recovers the well-known exponential decay of the structureless non-interacting Fermi gas bath \cite{Guinea85}. 

This non-trivial temperature dependence also allows to distinguish a non-interacting bath from the interacting SYK bath. The ancillary model itself does not lead to temperature-dependent impurity dynamics (see end of Sect.~V). The temperature dependence observed in Figs.~3b and~3c arises because the ancillary model itself changes as a function of time because $\rho_c(\omega)$ in eq. (\ref{rhocFG}) depends on temperature.

An obvious outlook is to consider interacting impurities coupled to a strongly interacting bath modeled as an SYK model. Work along these lines is in progress.\\
\section*{Data Availability} 
Simulation codes and the data shown in the figures are openly available on Zenodo \cite{zenodo}.

\section*{Acknowledgments}
We acknowledge prior collaboration and helpful discussions with A.~Simm that contributed to the motivation of our work.   
This work was funded by the Deutsche Forschungsgemeinschaft (DFG, German Research Foundation) – 499180199;
via FOR 5522, project~T1.

------------------------------------
\appendix
\section{Saddle point equations for impurity plus SYK bath setting}
Here we show the derivation of the Dyson equations for the system of impurity and SYK-4 bath in the large-$N$ limit. This can easily be generalized to arbitrary \(q\).  
We consider the Hamiltonian 
\begin{widetext}
\begin{equation}
    H = \epsilon_d d^\dagger d + \frac{1}{(2N)^{3/2}}  \sum^N_{1<i<j<k<l<N} J_{ijkl}c^\dagger_{i}c^\dagger_{j} c_{k} c_{l} + \frac{V}{\sqrt{N}}\sum\limits_{i=1}^N  \left(c_i^{\dagger} d+ d^{\dagger} c_i\right) , 
\end{equation}
\end{widetext}
for which the partition function averaged over random interactions \(J_{ijkl}\) is given by 
\begin{equation}
    Z = \int  \mathcal{D}(c^\dagger, c, d^\dagger, d) e^{-(S_{\text{imp}}+S_{\text{SYK}}+S_{\text{int}})} \ .
\end{equation}
The total action consists of impurity part
\begin{align}
    S_{\text{imp}} &= \int d\tau d\tau' d^\dagger(\tau)\left(\partial_\tau + \epsilon_d\right)\delta(\tau-\tau')  d(\tau'),
\end{align}
the SYK action \cite{PhysRevX.5.041025}
\begin{equation}
    S_{\text{SYK}} = \int d\tau  \sum_i c_i^\dagger \partial_\tau c_i - \frac{J^2}{4N^3}\iint d\tau d\tau' \left|\sum_i c_i^\dagger(\tau) c_i(\tau') \right|^4,
\end{equation}
and the interaction part
\begin{align}
    S_{\text{int}} = \int d\tau \frac{V}{\sqrt{N}}\sum_i (c_i^\dagger(\tau)d(\tau) + d^\dagger(\tau) c_i(\tau)).
\end{align}
After integrating out the bath, the partition function becomes
\begin{align}
Z = \int  \mathcal{D}( d^\dagger, d) e^{-S_{\text{imp}}} Z_{\text{bath}}[J]  = \int  \mathcal{D}( d^\dagger, d)e^{-S_\text{eff}}
\end{align}
where we defined the partition function of the bath as \(Z_{\text{bath}}[J]\), that depends on the source \(J = \frac{V}{\sqrt{N}} d\). The effective action is defined as \(S_\text{eff} = S_\text{imp} - \ln Z_\text{bath}[J]\), where the second term is a generating functional for connected bath correlation functions. Since only the two-point function contributes to the resonant level model dynamics, we can expand the second term of the effective action up to a leading term, meaning 
\begin{align} \label{expansion}
    \ln Z_\text{bath} = \ln Z[0] + \int d \tau d\tau'  J^\dagger(\tau) G_c(\tau-\tau') J(\tau') + \dots 
\end{align}
where the bath's two-point function is \begin{equation}\label{green's f}
    G_c(\tau', \tau) = -\frac{1}{N} \sum_i c_i(\tau') c_i^\dagger(\tau) .
\end{equation} 
From the expansion in eq. (\ref{expansion}), we can read off the self-energy. Combining it with the Dyson equation for the Green's function, we get 
\begin{align}
        G_d(i\omega_n) &= \frac{1}{i\omega_n +\epsilon_d- \Sigma_d(i\omega_n)} \\
    \Sigma_d(\tau) &= V^2 G_c(\tau)
\end{align}
In order to obtain the bath part of the Dyson equations, we follow the usual procedure by rewriting the impurity action as
\begin{align}
    S_{\text{imp}} &= \int d\tau \left[d^\dagger\left(\partial_\tau + \epsilon_d\right) d + \frac{V}{\sqrt{N}}\sum_i (c_i^\dagger d + d^\dagger c_i)\right] \nonumber \\&= \int d \tau \left(\tilde{d}^\dagger (G_d^0)^{-1} \tilde{d} - \frac{V^2}{N}\sum_{ij}c_i^\dagger G^0_d c_j\right).
\end{align}
Here we decoupled the impurity terms from the bath by introducing a new term \(\tilde{d} = d + G_d^0\,
\frac{V}{\sqrt{N}}\sum_i c_i\), where we define \((G_d^0)^{-1}=\partial_\tau+\epsilon_d\). The first term in the decoupled equation does not depend on $N$ and thus has no influence on the large-$N$ limit. Consequently, only the second term of \(S_{\text{imp}}\) contributes to the the total action. We introduce bilocal fields  \(G(\tau, \tau')\) and \(\Sigma(\tau, \tau')\) to the partition function by inserting an identity
\begin{eqnarray}
    1 = \int \mathcal{D} (G,\Sigma )\exp \bigg[ -\iint d\tau d\tau' \Sigma(\tau, \tau')  \nonumber  \\ \times \bigg(\sum_i c_i(\tau') c_i^\dagger(\tau) + N G(\tau',\tau) \bigg) \bigg] 
\end{eqnarray}
which should satisfy the constraint given by eq. (\ref{green's f}).
Combining the SYK and the impurity action with the above identity, we get an action 
\begin{widetext}
\begin{align}
    S &= \iint d\tau d\tau' \left\{\sum_i c_i^\dagger(\tau) \left(\delta(\tau-\tau')  \delta_{ij}\partial_{\tau'} -\frac{V^2}{N}G^0_d - \delta_{ij}\Sigma_c(\tau, \tau')\right)c_j(\tau') - \frac{J^2N}{4}\left|G_c(\tau, \tau')\right|^4+NG_c(\tau',\tau) \Sigma_c(\tau ,\tau') \right\} \\
     &= -N \left \{\log \left(\det\Big(\delta(\tau-\tau')\partial_{\tau'} -\frac{V^2}{N}G^0_d - \Sigma_c(\tau, \tau')\Big)\right) + \iint d\tau d\tau'\left(-\frac{J^2}{4}\left|G_c(\tau, \tau')\right|^4+G_c(\tau',\tau) \Sigma_c(\tau ,\tau')\right) \right\} 
\end{align}
\end{widetext}
where, on the second line, we integrated out the fermionic fields. Varying the action with respect to \(G\) and \(\Sigma\), we obtain the Schwinger-Dyson equations for the SYK bath
\begin{subequations}
\begin{align}
\frac{\delta S}{\delta G_c} &= 0 \quad \to \quad \Sigma_c(\tau) = -J^2 G_c^2(\tau) G_c(-\tau)
\\
\frac{\delta S}{\delta \Sigma_c} &= 0 \quad \to \quad G_c^{-1}(\tau)= \partial_\tau -\Sigma_c(\tau) - \frac{V^2}{N}G^0_d(\tau) 
\end{align}
\end{subequations}
These equations can be generalized to SYK-q in a following way
\begin{subequations}
\begin{align}
    G_c(i\omega_n) &= \frac{1}{i\omega_n - \Sigma_c(i\omega_n)} \\
    \Sigma_c(\tau) &= -J^2G_c^{q/2}(\tau)G_c^{q/2-1}(-\tau) + \frac{1}{N}V^2G^0_d(\tau) \\
    G_d(i\omega_n) &= \frac{1}{i\omega_n +\epsilon_d- \Sigma_d(i\omega_n)} \\
    \Sigma_d(\tau) &= V^2 G_c(\tau).
\end{align}
\end{subequations}

\bibliography{references.bib}
\end{document}